\documentclass[twocolumn,prl]{revtex4}
\usepackage{graphicx,bm,amsmath,amssymb}

\begin{document}

\title{Quantum interference in coherent molecular conductance}
%\title{Tunable quantum conductance through single molecules}
\author{Juli\'{a}n~Rinc\'{o}n, K. Hallberg}
\affiliation{Centro At\'{o}mico Bariloche and 
Instituto Balseiro, Comisi\'{o}n Nacional
de Energ\'{\i}a At\'{o}mica and CONICET, 8400 Bariloche, Argentina}

\author{S. Ramasesha}
\affiliation{Solid State and Structural Chemistry Unit,
Indian Institute of Science, Bangalore 560 012, India}

\date{\today}

\begin{abstract}
Coherent electronic transport through individual molecules is crucially 
sensitive to quantum interference. Using exact diagonalization techniques, 
we investigate the zero-bias and zero-temperature conductance through $\pi$-conjugated annulene
molecules (modeled by the 
Pariser-Parr-Pople and Hubbard Hamiltonians)
weakly coupled to two leads. We analyze the conductance for different 
source-drain configurations, finding an important reduction for certain 
transmission channels and for particular geometries as a consequence of
destructive quantum interference between states with definite momenta. 
When translational symmetry is broken by an external perturbation we find 
an abrupt increase of the conductance through those channels. Previous 
studies concentrated on the effect at the Fermi energy, where this effect 
is very small. By analysing the effect of symmetry breaking on the
main transmission channels we find a much larger response thus leading to 
the possibility of a larger switching of the conductance through single 
molecules.

\end{abstract}

\maketitle

The possibility of achieving controlled quantum transport through single 
molecules has become a reality as seen from various successful attempts 
in different systems\cite{aviram,reed,cuniberti}. 
The electronic transport through single $\pi$-conjugated 
molecules has been studied in several theoretical \cite{teo1,
teo2,teo4,teo5,teo11} and experimental\cite{
exp2,exp14,exp16,exp17} works. This important step leads not only to miniaturization 
of electronic devices but also 
to the possibility of taking advantage of new 
physical properties in these systems.

A fundamental physical property in quantum transport is quantum interference, 
which could be a handle  to control 
conductance through such 
systems. Constructive and destructive interference play a crucial role which, 
for most cases, is 
\emph{per se} independent of the structure and 
composition of the molecular bridges to the leads.

Even though changing from constructive to destructive interference might be
possible for certain geometries and molecules, disrupting the destructive
interference by perturbing the molecule seems to be more 
dramatic. This has
been proposed in two recent theoretical works \cite{teo4,teo5} for the
so-called Quantum Interference Effect Transistors (QuIET) based on
single annulene molecules, including benzene.
In the first paper, the systems are modeled by a 
Pariser-Parr-Pople (PPP) Hamiltonian\cite{ppp} using the Ohno
parametrization \cite{ohno} and solved
using the self-consistent Hartree-Fock (HF) approximation, 
while the second work resorts to ab-initio calculations 
within the LDA and ab-initio approximations. The conductance 
in both cases is then calculated for the strong coupling limit   
using the non-equilibrium 
Green's function\cite{hirose} and the Landauer-B\"uttiker formalism\cite{LB}.
Both studies concentrate on the equilibrium conductance at 
zero bias and gate voltage ($V_g=0$, {\it i.e.} the off-resonance 
tunnelling regime) which coincides with the Fermi energy of 
the leads. 
However, annulenes have a gap 
at this energy  due mainly to the strong Coulomb interactions present in the molecule
($N=4n+2$ annulenes like benzene, are closed-shell molecules with no level at zero 
energy even in the absence of interactions).
The conductance at zero gate voltage is finite, albeit small, only for strong coupling 
to the leads\cite{note}. For weak coupling the conductance will be zero and will be 
appreciable only through the main channels of the 
molecule which are a few eV away from the Fermi energy. By analysing the QuIET at the main 
transmittance channels the switching effect will be much more pronounced and robust as we 
will show below.

In this work we analyze the linear conductance of a series of
annulenes at finite gate voltages and several source-drain lead 
configurations. For leads set up at 180 degrees (the A configuration 
in Fig. 1), the system presents completely constructive 
interference in all transmission channels. 
However, for other lead
configurations, we find different behaviours depending on $N$. 
For $N=4n+2$ annulenes we 
find  a reduced conductance through particular channels due to 
a nearly destructive interferences explained below. When the 
molecular translational invariance is broken by means of elastic 
scattering or decoherence factors, we find an abrupt increase 
in the conductance through these channels due to disruption 
of destructive interference. This increase is much larger than 
was found in previous works since the focus is set on a resonance 
transmission channel through the molecule.

However, for $N=4n$ annulenes the effect is 
even more striking: shifting the drain lead by one lattice site 
from the A to the B configuration (see Fig. 1) reduces the 
transmission to zero. This is, in fact due completely to the destructive 
interference for $N=4n$ molecules as explained below. A translational 
symmetry breaking perturbation will have, 
as in this case, a huge effect, since the conductance through 
these channels will increase dramatically from zero to an appreciable
value. Hence, such molecules could have important applications. 
Previous theoretical studies analyze the effect of electron correlation on the structure 
stability of carbon rings exhibiting competing many-body effects of aromaticity, 
Jahn-Teller 
distortion and dimerization\cite{torelli,sondheimer}. These studies 
indicate that, under
certain conditions, some of the systems considered below could be stabilized.
In addition, ring geometries of quantum dots are also 
potential devices where such a behaviour could be observed.

\begin{figure}[tbp]
\includegraphics[width=7cm]{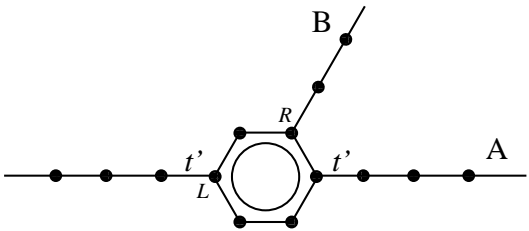}
\caption{Schematic representation of the molecules studied numerically 
in the presence of a gate voltage $V_g$, small hybridization to the 
leads $t'$ and two source-drain lead geometries.}
\label{molecules}
\end{figure}

In this work we consider an annular 
$\pi$-conjugated molecules with $N$ sites, weakly connected to
non-interacting leads in the A or B configurations (Fig.~\ref{molecules}).
We consider the following Hamiltonian for the system consisting of two 
leads weakly coupled to the molecule:
\begin{equation}
H=H_{\mathrm{ring}}+H_{\mathrm{leads}}+H_{\mathrm{links}}.  \label{esh}
\end{equation}%
The first term describes the 
isolated $\pi$-conjugated molecule, modeled by the PPP
Hamiltonian\cite{ppp}, with on-site energy given by a gate 
voltage $V_{g}$:
\begin{eqnarray}
H_{\mathrm{ring}}&=&- t \sum_{\langle ij\rangle,\sigma}   c_{i\sigma }^{\dag }  c_{j\sigma }
 + \sum_i U_i\left( n_{i\downarrow}-\frac{1}{2}\right)\left( n_{i\uparrow}-\frac{1}{2}\right) 
\nonumber \\
&&-eV_{g}\sum_{i,\sigma }  c_{i\sigma }^{\dag }  c_{i\sigma}
 + \sum_{i>j} V_{ij}\left( n_i-1\right) \left( n_j -1\right)
\end{eqnarray}
%\begin{eqnarray}
%\hat H_{\mathrm{ring}} =-eV_{g}\sum_{i=1,\sigma } ^N \hat c_{i\sigma }^{\dag } 
%\hat c_{i\sigma
%}- t \sum_{\langle ij\rangle,\sigma}  \hat c_{i\sigma }^{\dag } \hat c_{j\sigma }~ 
%\nonumber \\
% + \sum_i U_i~(\hat n_{i\downarrow}-1/2)(\hat n_{i\uparrow}-1/2) 
% + \sum_{i>j} V_{ij}(\hat n_i-1) (\hat n_j -1)
%\end{eqnarray}
where the operators $c_{i\sigma }^{\dag }$ ($c_{i\sigma }$) 
create (annihilate) an electron of spin $\sigma$ in the $\pi$ orbital 
of the Carbon atom at site $i$, $n$ are the corresponding number 
operators and $\langle\cdots\rangle$ stands for bonded
pairs of Carbon atoms.
The intersite interaction
potential $V_{ij}$ is parametrized so as to interpolate between $U$ 
and $e^2/r_{ij}$ in the limit $r_{ij} \longrightarrow \infty$ 
\cite{ohno}.  In the Ohno interpolation,  $V_{ij}$
is given by
\begin{equation}
V_{ij} \! = \! U_i(1+0.6117~r_{ij}^2)^{-1/2}
\end{equation}
where the distance $r_{ij}$ is in \AA. The standard Hubbard parameter 
for $sp^2$ Carbon is $U_i=11.26~\textrm{eV}$ and hopping parameter $t$
for $r=1.397~\textrm{\AA}$ is $2.4~\textrm{eV}$ \cite{rama}, and all energies 
are in eV. The second term in Eq.(1) corresponds to two 
tight-binding semi-infinite chains for the left and right leads. 
The third term in Eq.~(\ref{esh}) describes the coupling of the 
edge sites of
the left and right leads with sites $L$ (left) and $R$ (right) of the 
system respectively.
When the ground state of the isolated ring is non-degenerate, and the coupling
$t^{\prime }$ between the leads and the ring is weak ($t^{\prime}\ll t$), 
equilibrium conductance at 0 K
can be expressed to second order in $%
t^{\prime }$ in terms of the retarded Green's function for the isolated ring
between sites $i$ and $j$: $G_{i,j}^{\mathrm{R}}(\omega )$.
For an incident particle with energy $\omega =-2t\cos k$ and momentum $\pm k$%
, the transmittance reads \cite{Jagla,ihm}:
\begin{equation}
T(\omega ,V_{g})=\frac{4t^{2}\sin ^{2}k|{\tilde{t}}(\omega )|^{2}}{%
\left\vert [\omega -{\epsilon_1 }+te^{ik}][\omega -{\epsilon_2 }+te^{ik}]-
|{\tilde{t}}^{2}|\right\vert ^{2}},  \label{tra}
\end{equation}%
where
$\epsilon_{1(2)} (\omega )=t^{\prime \,2}G_{LL(RR)}^{\mathrm{R}}(\omega ),$ $
\tilde{%
t}(\omega )=t^{\prime \,2}G_{L R}^{\mathrm{R}}(\omega ),  \label{et}$
play the role of a correction to the on-site energy at the extremes of the
leads and an effective hopping between them respectively. The conductance is 
$G=(2e^{2}/h)T(\mu ,V_{g} )$, where $\mu $ is the Fermi level, which we set 
to zero (half-filled leads) and $V_g$ enters implicitly through the Green's 
functions.

In our study, the interacting system with $N$ sites is solved exactly 
using numerical techniques (Lanczos or Davidson)
to diagonalize huge matrices, and the Green's functions are obtained 
straightforwardly. The correlations are, thus, treated in an exact manner.

We now analyze the two possible scenarios for annulenes with $N=4n+2$ and $N=4n$
with two terminals. For the 
first case, the allowed total momentum quantum numbers are 
$k=2r\pi/(4n+2)=r\pi/(2n+1)$, 
with $r$ an integer, while for the second 
case the allowed momenta are $k=2r\pi/4n=r\pi/2n$.
For a two-terminal set up, wave functions travelling through both branches 
of the molecule will interfere producing different interference patterns 
depending on the positions of the leads. The phase difference will be momentum 
times the difference in the lenghts of the two trajectories (in units of the C-C 
separation): $\Delta \phi=k\Delta x$.

\begin{figure}[tbp]
%\vspace{1cm}
\includegraphics[width=7.5cm]{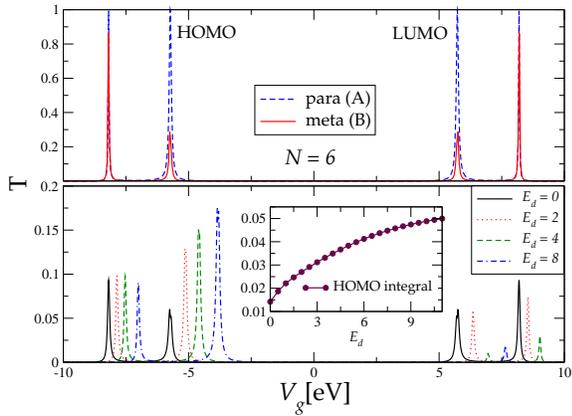}
\caption{Top: Transmittance through a benzene molecule for the ``para" (A 
substitution in Fig. 1) (broken line) and ``meta" (B substitution in Fig. 1) 
(full line) configurations as a function of the gate voltage measured from 
the Fermi energy. Here $t'=0.4$.  Bottom: Same
for the ``meta" configuration in the presence of different on-site potentials 
which break the translational invariance. Here a finite Lorentzian width $\eta=0.03$ and 
$t'=0.1$ have been taken for visualization purposes. Inset: evolution of 
the area under the transmission peak through the HOMO as a function of 
local energy.} 
\label{benzene}
\end{figure}

Specializing to benzene ($N=6$), for leads in the ``para" position,
$\Delta x=0$ and the waves are in phase, interfering constructively (Fig.~\ref{benzene}). However,
in the ``meta" position ($\Delta x=2$), the interference will depend on the 
$k$ value of the particular channel. For the 
highest occupied molecular orbital (HOMO) and lowest unoccupied molecular orbital (LUMO)
the phase differences will be $\Delta \phi=2\pi/3$ and  $\Delta \phi=4\pi/3$ respectively and the 
interference will reduce the amplitude in these
channels. For benzene configuration of the leads producing a completely destructive 
scenario for these channels
does not exist. Previous experimental and theoretical works 
\cite{teo14,teo16} noted that the transmission through benzene in the ``meta" 
configuration is much lower than that through the ``para" position. 

There 
is no transmission channel at $V_g=0$ since the density of states of the bare 
molecule at that energy is zero. A detailed analysis of quantum interference 
effects in benzene was carried out for the large coupling regime 
\cite{solomon} where interesting effects appear at zero energy and 
finite bias [\onlinecite{begemann,hettler}] 
at weak coupling.

What would happen if the translational symmetry is broken by an external 
perturbation? This question was first addressed in [\onlinecite{teo4}] by introducing 
a local energy ($\Sigma$), at one site in benzene, the real part of $\Sigma$
would produce elastic scattering and its imaginary part, decoherence. In that 
study, the focus was at
the Fermi energy of the leads (set to zero) where the observed effect is small.
However, by studying the effect of external perturbations on the 
main transmittance channels such as the HOMO and LUMO, we find a much larger
response. We show these results in the bottom part of figure 2, where an 
additional diagonal energy is added to the site to the right of the B (``meta") 
position (the effect is not qualitatively dependent on this position). It is 
clearly seen that, in this case, the small peak corresponding to the HOMO
level develops and grows as the local energy is increased, disrupting the 
translational symmetry responsible for the 
destructive interference (see inset).

For larger $4n+2$ rings, the momentum of the HOMO level approaches $\pi/2$ 
and the interference becomes more destructive. This is shown for a 
10 site annulene 
in figure 3, where the peaks for the B position of the leads are much smaller 
than those corresponding to benzene (Fig. 2, top). 
Growth of the HOMO 
peak with a local potential energy is larger in this case than in benzene.

\begin{figure}[tbp]
% \vspace{1cm}
\includegraphics[width=7.5cm]{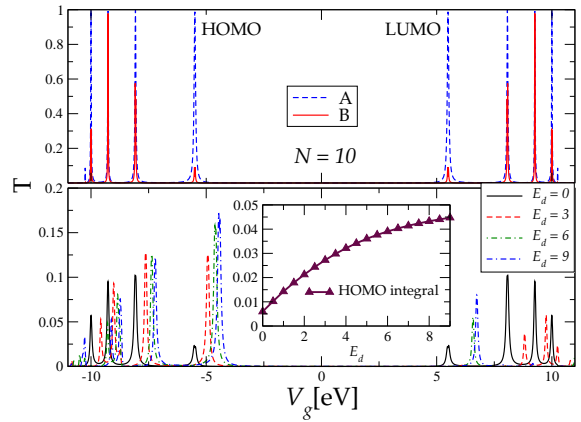}
\caption{Same as figure 2, for $N=10$ annulene.
}
%\label{10}
\end{figure}
However, for annulenes with 4n atoms, depending on the total
momentum of the particular channel, the interference can be totally
destructive. For these 
annulenes the interference will be completely
constructive in the A configuration
and totally destructive if one lead is shifted by one site (B configuration, 
see Fig. 1). In these cases we can observe the emergence
of a transmission channel, if the translational invariance is disrupted.

\begin{figure}[b]
\vspace{0.5cm}
\includegraphics[width=7.5cm]{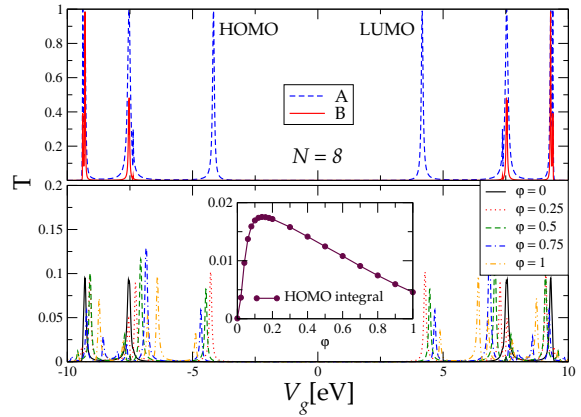}
\caption{Top: Transmittance through an [8]-annulene molecule for the A 
(broken line) and B (full line) configurations as a function of the gate 
voltage measured from the Fermi energy. Here $t'=0.4$. Bottom: 
Transmittance for the B configuration in the presence of a local 
perturbation of the hopping of the form $t(1-\varphi)$ which breaks 
the translational invariance ($\eta=0.03$ and $t'=0.1$). Inset: evolution of the HOMO integral as a function of $\varphi$.} \label{8}
\end{figure}

The effect of symmetry breaking by a local external perturbation is much 
more striking in annulenes with $4n$ sites. As discussed above, we expect 
a new peak to emerge when the translational symmetry is broken. In 
figure 4 we show results for conductance through an 8-site annulene 
represented by the PPP model. In the top figure, the HOMO and LUMO 
channels, fully formed for the A configuration, completely vanish
in the B case due to quantum interference. When a local perturbation is 
applied to the molecule, in this case a different hopping 
on the bond to the right of the B lead, $t(1-\varphi)$, the absent peak emerges 
as shown in the inset of the bottom figure. A similar effect occurs also 
for a local potential 

Finally, in figure 5 we present results for a [12]-annulene. In this case 
also the HOMO and LUMO peaks for the B configuration are non-existent due 
to destructive interference. Here we show how the HOMO peak arises as a 
function of a different hopping parameter in one lead of the form 
$t(1-\varphi)$ (top left) and as a function of dimerization, $\delta$,
for transfer integral for
$t(1-(-1)^i\delta)$ of the bond between sites $i$ and $i+1$ (top right). 
The transmittance in the A configuration doesn't change in the presence 
of dimerization, while in case B,
the HOMO and LUMO channels develop 
suddenly already for a small $\delta$ (bottom panel).
Dimerization of the annulenes lifts the two-fold degeneracy of the HOMOs 
as well as the LUMOs in large annulenes. Thus, the condition for destructive
interference between the two paths is lifted and one observes transmission
through these channels.

In conclusion, by analysing the resonant conductance through 
the HOMO and LUMO channels in 
the weak lead-molecule coupling regime we find a 
strong dependence on the source-drain configuration and on the molecular geometry due to 
quantum interference. 
This effect is more robust and striking than in the strong-coupling case since in the latter 
the destructive interference occurs only at zero gate voltage and 
can be masked in an experiment, 
while in the former a whole channel can appear or disappear depending on the 
geometry. We'd also like to stress that these results are not affected by molecular 
vibrations at room temperature since modes that can cause decoherence are excited at 
temperatures higher than $500K$\cite{teo4}.

\begin{figure}[tbp]
% \vspace{1cm}
\includegraphics[width=7.5cm]{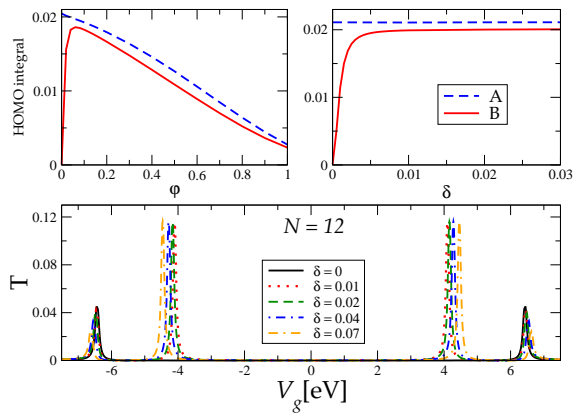}
\caption{Top: HOMO integral of a [12]-annulene molecule for the A (broken line) 
and B (full line) configurations as a function of the locally perturbed hopping 
$t(1-\varphi)$ (left panel) and the dimerization parameter $\delta$ (right panel).
Bottom: Transmittance for the B configuration in the presence of dimerization.}
\label{12}
\end{figure}

In [$4n+2$]-annulenes like benzene in 
the B (``meta'') configuration (Fig. 1) the interference will be nearly destructive. 
The larger the molecule the closer 
the phase difference between different paths is to $\pi$ and hence 
the interference becomes more destructive. For 
[$4n$]-annulenes, on the other hand, the interference for these configurations 
can be completely destructive. We have shown the effect of 
symmetry-breaking 
perturbations on different transmission channels. For 
[$4n$]-annulenes the effect is striking since the transmission changes from 
zero to a 
large finite value for small perturbations. These effects 
should be seen in substituted annulenes and should also be appreciable in
transport through rings of quantum dots.

We thank A. A. Aligia and B. Alascio for useful discussions. This work 
was done in the framework of projects PIP 5254 of CONICET, PICT 2006/483 
of the ANPCyT and the ARG/RPO-041/2006 Indo-Argentine collaboration. 
SR thanks DST for support through JC Bose fellowship.


\begin{thebibliography}{99}

\bibitem{aviram} A. Aviram, M. A. Ratner, Chem. Phys. Lett. \textbf{29}, 277 (1974).

\bibitem{reed} M. A. Reed, {\it et al.}, Science \textbf{278}, 252 (1997).

\bibitem{cuniberti} G. Cuniberti, G. Fagas, K. Richter (eds), Introducing Molecular Electronics, Springer, Berlin \textbf{2005}.

\bibitem{teo1} A. Nitzan, M. A. Ratner, Science \textbf{300}, 1384 (2003).

\bibitem{teo2} N. Tao, Nature Nanotech. \textbf{1}, 173-181 (2006).

\bibitem{teo4} D. Cardamone, {\it et al.}, 
%C. Stafford, S. Mazumdar, 
Nano Lett. \textbf{6}, 2422 (2006).

\bibitem{teo5} S-H. Ke, 
%W. Yang, H. Baranger, 
{\it et al.}, Nano Lett. \textbf{8}, 3257 (2008).

\bibitem{teo11} S. Yeganeh, 
%M. Ratner, M. Galperin, A. Nitzan, 
{\it et al.}, Nano Lett. \textbf{9}, 1770 (2009).

\bibitem{exp2} J. Park, {\it et al.}, Nature \textbf{417}, 722-725 (2002).

\bibitem{exp14} B. Venkataraman, {\it et al.}, Nature \textbf{442}, 904 (2006).

\bibitem{exp16} A. V. Danilov, {\it et al.}, Nano Lett. \textbf{8}, 1 (2008).

\bibitem{exp17} T. Dadosh, {\it et al.}, Nature \textbf{436}, 677 (2005).

\bibitem{ppp} R. Pariser, R. Parr, J. Chem. Phys. \textbf{21}, 466 (1953); J. A. Pople, Trans. of the Faraday Soc.,  \textbf{49}, 1375 (1953).

\bibitem{ohno} K. Ohno, Theor. Chim. Acta \textbf{2}, 219 (1964).

\bibitem{hirose} K. Hirose, "First-principles calculations in real space 
formalism: Electronic configurations and transport properties of nanostructures", 
Imperial College, London \textbf{2005}.

\bibitem{LB} M. B\"uttiker, 
%Y. Imry, R. Landauer, S. Pinhas, 
{\it et al.}, Phys. Rev. B \textbf{31}, 6207 (1985).

\bibitem{note} For certain geometries ({\it e.g.} the meta configuration) 
the conductance at $V_g=0$ for the strong coupling regime can be zero due 
to quantum interference effects\cite{solomon}, but not between paths with 
$k=\pi/2$ as stated in [\onlinecite{teo4,teo5}], since this is not an allowed 
quantum number for benzene molecules.

\bibitem{torelli} T. Torelli, L. Mitas, Phys. Rev. Lett. \textbf{85}, 1702 (2000).

\bibitem{sondheimer} F. Sondheimer, Acc. Chem. Res. \textbf{5}, 81 (1972).

\bibitem{rama} Z. G. Soos, and S. Ramasesha, Phys. Rev. B \textbf{29}, 5410 (1984).

\bibitem{Jagla} E. Jagla, C. Balseiro, Phys. Rev. Lett. \textbf{70}, 639 (1993).

\bibitem{ihm} A. A. Aligia, 
%K. Hallberg, B. Normand, A. Kampf, 
{\it et al.}, Phys. Rev. Lett. \textbf{93}, 076801 (2004).

\bibitem{teo14} S. N. Yaliraki and M. A. Ratner, Ann. N. Y. Acad. Sci. \textbf{960}, 
153 (2002).

\bibitem{teo16} D. Walter, 
%D. Neuhauser, and R. Baer, 
{\it et al.}, Chem. Phys. \textbf{299}, 139 (2004).

\bibitem{solomon} G. Solomon, {\it et al.}, Chem. Phys. \textbf{129}, 054701 (2008).

\bibitem{begemann} G. Begemann, 
%D. Darau, A. Donarini M. Grifoni, 
{\it et al.}, Phys. Rev. B \textbf{77}, 201406(R) (2008); \textbf{78}, 089901(E) 
(2008).

\bibitem{hettler} M. H. Hettler, 
%W. Wenzel, M. R. Wegewijs, H. Schoeller, 
{\it et al.}, Phys. Rev. Lett. \textbf{90}, 076805 (2003).

\end{thebibliography}
\end{document}